\input{aipcheck}

\documentclass[
    ,final            
  ]
  {aipproc}

\layoutstyle{8x11single}

\begin{document}

\title{The Formation of Low-Mass Double White Dwarfs through an Initial Phase of Stable Non-Conservative Mass Transfer}


\author{T. E. Woods}{
  address={University of Alberta, 1322 89 Ave, Edmonton, AB, T6G 2E7, Canada},
  email={tewoods@ualberta.ca},
  thanks={This work was commissioned by the AIP}
}

\author{N. Ivanova}{
  address={University of Alberta, 1322 89 Ave, Edmonton, AB, T6G 2E7, Canada},
  email={nata.ivanova@ualberta.ca},
}

\author{M. van der Sluys}{
  address={University of Alberta, 1322 89 Ave, Edmonton, AB, T6G 2E7, Canada},
  altaddress={Radboud University Nijmegen, P.O. Box 9010 NL-6500 GL Nijmegen, The Netherlands },
  email={m.vdsluys@ualberta.ca},
  homepage={http://www.astro.ru.nl/~sluys/}
}

\author{S. Chaichenets}{
  address={University of Alberta, 1322 89 Ave, Edmonton, AB, T6G 2E7, Canada},
  email={},
}
\begin{abstract}
 Although many double white dwarfs (DWDs) have been observed, the evolutionary channel by which they 
  are formed from low-mass/long-period red-giant--main-sequence (RG-MS) binaries remains uncertain. 
  The canonical explanations involve some variant of double common-envelope (CE) evolution, however it has been found that such a mechanism cannot produce the observed distribution. 
  We present a model for the initial episode of mass transfer (MT) in RG-MS binaries, and demonstrate that their evolution 
  into double white dwarfs need not arise through a double-CE process, as long as the initial primary's core mass ($M_{\rm{d,c}}$) does not exceed $0.46 M_{\odot}$. Instead, the first episode of dramatic mass loss may be stable, 
  non-conservative MT. 
  We find a lower bound on the fraction of transferred mass that must be lost from the system in order to provide for MT, and demonstrate the feasibility of this channel in producing observed low-mass (with $M_{\rm{d,c}} < 0.46 M_{\odot}$) DWD systems. 
\end{abstract}

\maketitle


In a CE phase, both components of a binary become engulfed in the envelope of the more evolved star. The envelope is then expelled at the cost of dramatically shrinking the orbit \cite{Pa76, Webbink84}. It has been shown \cite{Nelemans00, Sluys2006} that consideration of two subsequent CE events in which each 
  phase obeys the standard $\alpha_\mathrm{CE}$-prescription (based on energy considerations, with $\alpha _{\mathrm{CE}}$ as an efficiency factor), as well as stable conservative MT followed by 
  a single such CE event, failed to explain the observed DWD systems, proving inconsistent with the observed distribution of periods and mass ratios (mean mass ratio approximately 1). An alternative description considering angular momentum balance (the $\gamma$-formalism) was proposed \cite{Nelemans05}; however, this does not allow for strong constraints on the evolution of the system \cite{Webbink08, Zorotovic10}, proving unstable against small perturbations \cite{Woods10}. We revisit the possibility that the first phase of MT occurs as stable, non-conservative Roche-lobe overflow (RLOF), and demonstrate
that this channel can provide an adequate model for a number of observed DWDs when considering progenitor systems in which both components
are approximately 1 - 1.3 $M_{\odot}$---significantly smaller than previously considered \cite{Han1998}. 

\subsection{Stability Criteria}

In order for MT to proceed in a stable fashion, we require that the radius of the donor shrinks faster due to mass loss than its Roche lobe does (that is, $\zeta_{\rm{RL}} \leq \zeta_{\rm{ad}}$ given $R \propto M^{\zeta}$). The response of the Roche lobe is given by:
\begin{center}
\begin{equation}
  \zeta_\mathrm{RL}  \equiv  \frac{d\log\,R_\mathrm{RL}}{d\log\,M_\mathrm{d}} = 
\frac{\partial \ln a}{\partial \ln M_\mathrm{d}} + \frac{\partial \ln (R_{L}/a)}{\partial \ln q}\frac{\partial \ln q}{\partial \ln M_\mathrm{d}}.
\end{equation}
\end{center}
\noindent (with donor mass $M_{\rm{d}}$, accretor mass $M_{\rm{a}}$, Roche radius $R_{\rm{RL}}$, and orbital separation $a$) where $\partial \ln (R_{\rm{RL}}/a)/\partial \ln q$ depends on the system geometry, and the other terms depend on the masses as well as the
conservation factor ($\beta$, the fraction of mass lost by $M_{\mathrm{d}}$ that is accreted onto $M_{\mathrm{a}}$). A low mass star on the red-giant branch (RGB) can be well 
modelled as a condensed polytrope, and from this we can approximate the adiabatic response of the donor's radius as a simple function of 
the core mass. This implies that for given system parameters ($M_{\rm{d}}$, $M_{\rm{a}}$, $M_{\mathrm{d,c}}$, and to some extent $P_{\rm{orb}}$) we have a maximum fraction of 
mass lost by the primary which can be accreted onto the secondary while maintaining stability. As $\zeta_{\rm{RL}}$ and $\zeta_{\rm{ad}}$ both depend monotonically on mass loss, we need only consider stability at the onset of MT. For this work we adopt $\beta = 0.3$,
as it provides for generally stable MT given the above condition and our initial parameters.  

\subsection{Subsequent Evolution}

Using a variant of Eggleton's detailed stellar-evolution code {\bf ev} (referred to as {\bf STARS}) \cite{Eggleton1971, EKE2002, Glebbeek08}, we evolve a set of 1.2+1.1 $M_{\odot}$ systems with a range of initial periods through RLOF 
of the primary on the RGB. Our simulations demonstrate that MT proceeds stably until the envelope has been almost entirely stripped. 
The subsequent RLOF of the (original) secondary can then be expected to take place in an unstable manner, as the mass ratio has been significantly increased. Given an ensuing CE (in which the central concentration of the donor envelope can be calculated from our stellar models) with $\alpha _{\mathrm{CE}} = 0.5$, we find that our results fit within observations nicely, however further studies will be needed in order to determine if we can fully reproduce the observed distribution (in particular, those systems with q < 1).  This, as well as higher mass systems, may require some intermediate mechanism in the initial phase, removing the majority of the envelope with little orbital decay, as suggested by \cite{Nelemans00, Nelemans05}. Nonetheless, the stable non-conservative MT + CE channel provides an elegant solution for an initial ``envelope-removal event'' with a modest increase in period.

\begin{figure}
  \includegraphics[height=.23\textheight]{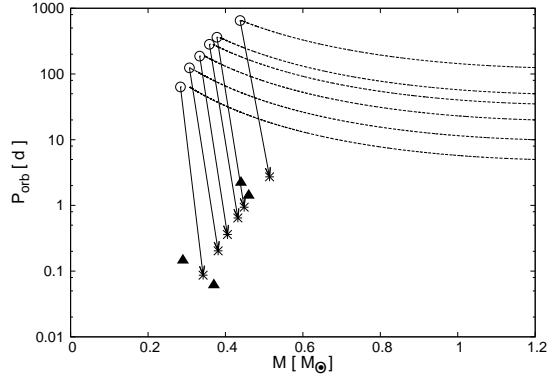}
  \caption{ Circles indicate $M_{\mathrm{d,c}}$ (taken as the total helium mass) and orbital period after the initial phase of stable MT.
  Stars indicate the final (core) mass of the secondary and orbital period after the second, unstable episode of mass loss (a CE-phase with $\alpha _{\mathrm{CE}} = 0.5$). 
  Solid lines connect the two relevant points for each system. Dotted lines mark evolution of total donor mass during initial MT phase. Triangles mark observed parameters (orbital period and mass of the
inferred secondary, taken from \cite{Sluys2006}).
\label{fig:endstate}}
\end{figure}

\begin{theacknowledgments}
  We thank P. P.\ Eggleton and E.\ Glebbeek for making their binary-evolution code available to us, as well as Gijs Nelemans for helpful discussion.  
  NI acknowledges support from the Canada Research Chairs and NSERC. MvdS acknowledges support from a CITA National Fellowship to the University of Alberta. 
\end{theacknowledgments}



\bibliographystyle{aipproc}   

\bibliography{references}

\IfFileExists{\jobname.bbl}{}
 {\typeout{}
  \typeout{******************************************}
  \typeout{** Please run "bibtex \jobname" to optain}
  \typeout{** the bibliography and then re-run LaTeX}
  \typeout{** twice to fix the references!}
  \typeout{******************************************}
  \typeout{}
 }

\end{document}